\documentclass[harvard,12pt]{iopart}
\usepackage{graphicx,harvard}

\usepackage{epsfig,amsbsy,bm,amssymb}
\usepackage[usenames,dvipsnames,svgnames,table]{xcolor}

\newcommand{\ba}{\begin{eqnarray}}
\newcommand{\ea}{\end{eqnarray}}
\renewcommand{\br}{\begin{eqnarray*}}
\newcommand{\er}{\end{eqnarray*}}
\newcommand{\be}{\begin{equation}}
\newcommand{\ee}{\end{equation}}
\newcommand{\bp}{\begin{minipage}}
\newcommand{\ep}{\end{minipage}}
\newcommand{\cn}{\citeasnoun}
  \newcommand{\mc}{\multicolumn}
\newcommand{\bt}{\begin{tabular}}
\newcommand{\et}{\end{tabular}}
\renewcommand{\l}{\lambda}

\newcommand{\f}{{\bm f}}
\renewcommand{\r}{{\bm r}}
\renewcommand{\k}{{\bm k}}

\newcommand{\z}{{\bm z}}
  
   \newcommand{\n}{{\bm n}}

\newcommand{\nn}{\nonumber}

\newcommand{\hs}{\hspace*}
\newcommand{\vs}{\vspace*}

\renewcommand{\bs}{\bigskip}

\renewcommand{\Eref}[1] {Eq.~(\ref{#1})}

  \newcommand{\la}{\langle}
  \newcommand{\ra}{\rangle}

\newcommand{\ket}[1]{| \: #1 \: \rangle}
\newcommand{\braket}[2]{\langle \: #1 \: | \: #2 \: \rangle}
\newcommand{\braOket}[3]{\langle \: #1 \: | \: #2 \:| \: #3 \: \rangle}

\newcommand{\intsum}[1]{\sum_{#1} \! \! \! \! \! \! \! \! \int }

\begin{document}

\title{On the angular dependence of the photoemission time delay in helium }

\author{I A Ivanov$^{1,2}$, J M Dahlstr\"{o}m$^{3}$, E
  Lindroth$^3$  and A~S~Kheifets$^1$}

\address{$^1$ Research School of Physics and Engineering, The
Australian National University, Canberra ACT 0200, Australia}

\address{$^2$ Center for Relativistic Laser Science, Institute for Basic Science (IBS), Gwangju 500-712, Republic of Korea} 

\address{$^3$ Department of Physics, Stockholm University, AlbaNova
University Center, SE-10691 Stockholm, Sweden}

\begin{abstract}

We investigate an angular dependence of the photoemission time delay
in helium as measured by the RABBITT (Reconstruction of Attosecond
Beating By Interference of Two-photon Transitions) technique. The
measured time delay 
$
\tau_a=\tau_W+\tau_{cc}
$
contains two distinct
components: the Wigner time delay $\tau_W$ and the continuum-continuum
(CC) correction $\tau_{cc}$.  In the case of helium with only one
$1s\to Ep$ photoemission channel, the Wigner time delay $\tau_W$
does not depend on the photoelectron detection angle relative to the
polarization vector. However, the CC correction $\tau_{cc}$ shows a
noticeable angular dependence. We illustrate these findings by
performing two sets of calculations. In the first set, we solve the
time-dependent Schr\"odinger equation for the helium atom ionized by
an attosecond pulse train and probed by an IR pulse. In the second
approach, we employ the lowest order perturbation theory which
describes absorption of the XUV and IR photons.  Both calculations
produce close results in a fair agreement with experiment.

\end{abstract}

\maketitle

\section{Introduction}

The generation and application of attosecond pulses through high-order
harmonic generation (HHG) has lead to experimental observation of
atomic processes taking place on an unprecedentedly short time scale
down to tens of attoseconds \cite{krausz:rmp:09}.  The two main
techniques, originally developed for characterization of attosecond
pulses, were employed in these studies. The RABBITT (Reconstruction of
Attosecond Beating By Interference of Two-photon Transitions)
technique \cite{MullerAPB2002,TomaJPB2002} was developed for
characterization of attosecond pulse trains (APT)
\cite{PaulScience2001}.  For single attosecond pulses (SAP)
\cite{HentschelNature2001}, the attosecond streak camera
\cite{ItataniPRL2002} and the FROG-CRAB technique (frequency-resolved
optical gating for complete reconstruction of attosecond bursts)
\cite{MairessePRA2005}, were developed.  The streak camera has been
applied to study electron dynamics in photoionization of solid
targets, such as tungsten \cite{cavalieri:07}, and the neon atom
\cite{SchultzeScience2010}.  In both cases, relative delays in
photoemission between different initial states have been reported.
Similarly, experiments based on the RABBITT technique with APTs have
demonstrated a relative time delay between photoemission from the $3s$
and $3p$ shells of argon
\cite{PhysRevLett.106.143002,PhysRevA.85.053424}.  In further
developments, the relative time delay between the outer shells of the
atomic pairs (He vs. Ne and Ne vs. Ar) has been determined owing to
active stabilization of the RABBITT spectrometer
\cite{0953-4075-47-24-245602}.  In conjunction with the HHG, the
RABBITT technique has also been used to determine the time delay in Ar
\cite{PhysRevLett.112.153001}.  Finally, measurement has been
performed in heavier noble gas atoms relative to the time delay in the
$1s$ sub-shell of He \cite{0953-4075-47-24-245003}. A recent review of
the field of attosecond chronoscopy of photoemission has been given by
\cn{RevModPhys.87.765}

Periodic trains of attosecond pulses typically consist of two pulses
of opposite polarities per fundamental laser cycle.  This translates to
a coherent comb of odd XUV harmonics, $\omega_{2q+1}=(2q+1)\omega$ of the
fundamental laser frequency $\omega$, in the frequency domain.  The
RABBITT technique builds on the interference of two ionization
processes leading to the same photoelectron state by (i) absorption of
$\omega_{2q-1}$ and $\omega$ or (ii) absorption of $\omega_{2q+1}$ and
stimulated emission of $\omega$.  In the following, we will label the
path with absorption of an IR photon by $(+)$ and that with emission
of an IR photon by $(-)$.  Both ionization processes lead to the
appearance of a side band (SB), in between the one-photon harmonic
peaks in the photoelectron spectrum, at the kinetic energy
$2q\omega-I_p$, where $I_p$ is the binding potential of the atom.  The
sideband magnitude oscillates with the relative phase between the
XUV and IR pulses \cite{MullerAPB2002,TomaJPB2002}
\be 
S_{2q}(\tau) = A+B\cos(2\omega \tau
-\Delta\phi_{2q}-\Delta\theta_{2q}) \ , 
\label{oscillation}
\ee
where $\tau=\varphi/\omega$ denotes the phase delay of the IR field. 
%
%
The term $\Delta\phi_{2q}=\phi_{2q+1}-\phi_{2q-1}$ denotes the phase difference between 
two neighbouring odd harmonics $2q\pm1$  that is related to the
finte-difference group delay of the attosecond pulse as 
$\tau_{2q}^{(\rm GD)}=\Delta\phi_{2q}/2\omega$.  
The quantity $\tau-\tau_{2q}^{(\rm GD)}$ is the delay 
between the maxima of the electric field laser oscillation 
and the group energy of the XUV pulse at $\omega_{2q}$.  
The additional term
$\Delta\theta_{2q}=\theta^{(-)}_{2q+1}-\theta^{(+)}_{2q-1}$ arises
from the phase difference of the atomic ionization amplitude for path
$(-)$ and $(+)$, respectively.  This phase difference can be converted
to the atomic delay
\be 
\tau_a = \Delta\theta_{2q}/2\omega,
\label{atomicdelay}
\ee
which can be interpreted as the sum of
the two distinct components \cite{Dahlstrom201353}
\be
\tau_a=\tau_W+\tau_{cc} \ .
\label{atomicdelayparts}
\ee
Here $\tau_W$ is the Wigner-like time delay associated with the XUV
absorption and $\tau_{cc}$ is a correction due to the
continuum--continuum (CC) transitions in the IR field.  The latter
term, $\tau_{cc}$, can also be understood as a coupling of the
long-range Coulomb ionic potential and the laser field in the context
of streaking \cite{PhysRevA.84.033401,C3FD00004D}.

The attosecond streak camera method 
is in many regards similar to the RABBITT method, 
the main difference being that `streaking' relies 
on isolated attosecond pulses corresponding to a 
continuum of XUV frequencies rather than the discrete
odd high-harmonics of the RABBITT method. 
The target electron is first ejected by the isolated XUV pulse 
and it is then streaked: accelerated or decelerated by the IR dressing field. 
In this technique, the photoelectron is detected in the direction of the joint
polarization axis of the XUV and IR fields. 
%
%
The RABBITT measurement is different in this respect because the
photoelectrons can be collected in {\it any} direction, in fact,
photoelectrons are often collected in all angles.  Hence a possible
angular dependence of the time delay may become an issue. Because of
the known propensity rule \cite{PhysRevA.32.617}, the XUV
photoionization transition $n_il_i\to El$ is dominated by a single
channel $l=l_i+1$. In this case, the Wigner time delay is simply the
energy derivative of the elastic scattering phase in this dominant
channel $\tau_W=d\delta_l/dE$. However, if the nominally stronger
channel goes through a Cooper minimum, the weaker channel with
$l=l_i-1$ becomes competitive. The interplay of these two
photoionization channels leads to a strong angular dependence of the
Wigner time delay because these channels are underpinned by different
spherical harmonics.  The hint of this dependence was indeed observed
in a joint experimental and theoretical study
\cite{0953-4075-47-24-245003} near the Cooper minimum in the $3p$
photoionization of argon.  This effect was seen as a much better
agreement of the angular averaged atomic calculations in comparison
with angular specific calculations. 
In subsequent theoretical studies, this effect was investigated in more
detail and an explicit angular dependence was graphically depicted
\cite{Dahlstrom2014,0953-4075-48-2-025602}.

In the case of a single atomic photoionization channel, like $1s\to
Ep$ channel in He, the interchannel competition is absent and the
Wigner time delay is angular independent. The early investigations of
the $\tau_{cc}$ correction \cite{Dahlstrom201353} showed 
no dependence of $\tau_{cc}$ over various angular momentum paths in
hydrogen, e.g. the ATI transitions
$s\rightarrow p\rightarrow s$  and $s\rightarrow p\rightarrow d$ 
showed $\tau_{cc}$ in excellent agreement. 
Hence one may
think that the RABBITT measured time delay in He should be angular
independent. This assumption was challenged in a recent experiment by
\cn{2015arXiv150308966H} in which the RABBITT technique was
supplemented with the COLTRIMS (Cold Target Recoil Ion Momentum
Spectroscopy) apparatus. This combination made it possible to relate
the time delay to a specific photoelectron detection angle relative to
the joint polarization axis of the XUV and IR pulses. The finding of
\cn{2015arXiv150308966H} is significant because the helium
atom is often used as a convenient reference to determine the time
delay in other target atoms. If the RABBITT measurement is not angular
resolved, like in the experiments by
\cn{0953-4075-47-24-245003} or \cn{0953-4075-47-24-245602},
the angular dependence of the time delay in the reference atom may
compromise the accuracy of the time delay determination in other
target atoms.
This consideration motivated us to investigate theoretically the
angular effects in the time delay of helium measured by the RABBITT
technique. 

The paper is organized as follows.
In Sec.~\ref{sec:theory} we present  our theoretical models. 
%
%
In Sec.~\ref{sec:theory:LOPT} we describe a frequency-domain method
based on the lowest-order perturbation theory (LOPT) for the
two-photon XUV and IR above-threshold ionization (ATI).  This
frequency-domain method is numerically efficient and allows for
inclusion of correlation effects by the many-body perturbation theory
(MBPT).
In Sec.~\ref{sec:theory:TDSE} we present a time-domain method based on
solution of the time-dependent Schr\"odinger equation (TDSE) within
the single-active electron approximation (SAE).  The helium atom is
subjected to the APT and the IR pulse and then, after the interaction
with the fields is over, the solution of the TDSE is projected on the
scattering states of the field-free Hamiltonian to extract the
angle-resolved photoelectron spectrum.
In Sec.~\ref{sec:results} we compare our results of the
frequency-domain and time-domain methods with recent experiments
\cn{2015arXiv150308966H}.  Finally, in Sec.~\ref{sec:conclusion} we
draw our conclusions.

\section{Theory and numerical implementation}
\label{sec:theory}
\subsection{LOPT approach}
\label{sec:theory:LOPT}

The RABBITT process can be described using the LOPT  with
respect to the dipole interaction with the XUV and IR fields.  The
dominant lowest-order contributions are given by two-photon matrix
elements from the initial electron state $i$ to the final state $f$ by
absorption of one XUV photon $\omega_x$, followed by exchange of one
IR photon $\omega$,
\be
M(f,\omega,\omega_x,i)=
\frac{1}{i}E(\omega) E(\omega_x)
\lim_{\varepsilon\rightarrow 0^+}
\intsum{p}
\frac{ \braOket{f}{z}{p} \braOket{p}{z}{i} }
{\epsilon_i+\omega_x-\epsilon_p+i\varepsilon},
\label{MqoOa}
\ee
where both fields are linearly polarized along the $\n_z$-axis.  The
single-electron states are expressed as partial wave states 
$\braket{\r}{i}=R_{n_i,\ell_i}(r)Y_{\ell_i,m_i}(\n_\r)$ and
$\braket{\r}{f}=R_{k_f,\ell_f}(r)Y_{\ell_f,m_f}(\n_\r)$ for bound
initial state and continuum final state with corresponding
single-particle energies $\epsilon_i<0$ and $\epsilon_f>0$, respectively.
Energy conservation of the process is given by
$\epsilon_f-\epsilon_i=\omega_x\pm\omega$, where $+(-)$ corresponds to
absorption (emission) of an IR photon.  All intermediate unoccupied
states, $\braket{\r}{p}=R_{n_p,\ell_p}(r)Y_{\ell_p,m_p}(\n_\r)$, are
included in the integral sum in Eq.(\ref{MqoOa}).  Angular momentum
conservation laws applied to the $1s^2$ ground state in helium require
that $\ell_i=0$, $\ell_p=1$ and $\ell_f=0,2$ and $m_i=m_p=m_f=0$.
The two-photon matrix element in Eq.~(\ref{MqoOa}) can be re-cast as a
one-photon matrix element between the final state and an uncorrelated
perturbed wave function (PWF)
\be
M(f,\omega,\omega_x,i)=\frac{1}{i}E(\omega_x)E(\omega) 
\braOket{f}{z}{\rho^{(0)}_{\omega_x,i}}.
\ee
The PWF is a complex function that describes the outgoing
photoelectron wave packet, with momentum $k'$ corresponding to the
on-shell energy $\epsilon'=\epsilon_i+\omega_x$, after absorption of
one XUV photon from the electron state $i$
\cite{Aymar1980,TomaJPB2002,Dahlstrom201353}.  Correlation effects due
to the screening by other electrons can be systematically included by
the MBPT, e.g. by substitution of the
uncorrelated PWF with the correlated PWF based on the Random-Phase
Approximation with Exchange (RPAE),
$\ket{\rho^{(0)}_{\omega_x,i}}\rightarrow\ket{\rho^{\rm (RPAE)}_{\omega_x,i}}$
\cite{Dahlstrom2014}.

\begin{figure}[ht]
\bp{0.5\textwidth}
\includegraphics[width=\textwidth]{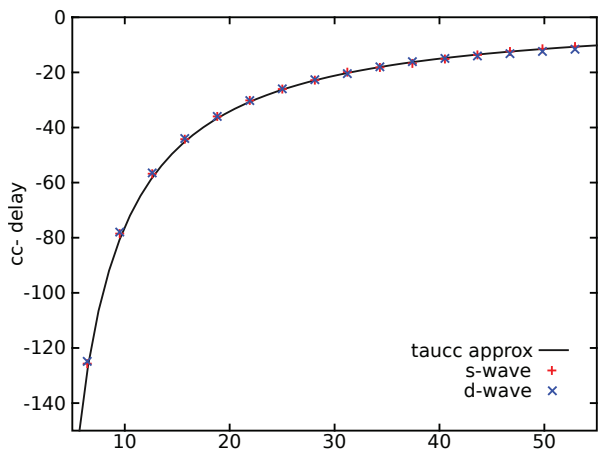}
\ep
\hs{-0.1\textwidth}
\bp{0.6\textwidth}
\caption{\label{Hetaucc} Continuum--continuum delay, $\tau_{cc}(l_f)$,
extracted from two-photon matrix elements for absorption and emission
of IR photons on particular partial wave paths leading to $l_f$ from
the helium ground state.  The black line the corresponding
$\tau_{cc}(l_f)$ of hydrogen \cite{Dahlstrom201353}.}  \ep
\end{figure}

For simplicity, we first consider a final state with angular momentum
$l_f$ that can be reached by two paths (i) absorption of two photons
$\omega_{2q-1}=(2q-1)\omega$ and $\omega$,
denoted $M_f^{\rm (+)}=M(f,\omega,\omega_{2q-1},i)$;
and (ii) absorption of one photon $\omega_{2q+1}=(2q+1)\omega$
followed by stimulated emission of $\omega$,
denoted $M_f^{\rm (-)}=M(f,-\omega,\omega_{2q+1},i)$.
The probability for detection of such an electron is proportional to 
\be 
S_{2q}(l_f)=\left|M_f^{\rm
(-)}\exp[i(\phi_{2q+1}-\varphi)]+M_f^{\rm
(+)}\exp[i(\phi_{2q-1}+\varphi)]\right|^2, 
\ee
where we write explicitly the phases of the fields,
$\varphi=\omega\tau$ for the $\omega$-field and $\phi_{2q+1}$ and
$\phi_{2q-1}$ for the $\omega_{2q+1}$ and $\omega_{2q-1}$-fields,
respectively.  The field amplitudes, $E$ inside $M$, are then real and
immaterial in this derivation.  The different signs of $\varphi$ in
the terms on the right side of Eq.(\ref{probq}) arise due to the IR
photon being either absorbed of emitted in the process,
$E(\omega)=|E(\omega)|e^{i\varphi}=E^*(-\omega)$.  Using
Eq.~(\ref{oscillation}) and Eq.~(\ref{atomicdelay}), the corresponding
atomic delay is
\be 
\tau_a(l_f)=\Delta\theta_{2q}(l_f)/2\omega= \arg\left[{M}^{(\rm
-)}_{f}{ M}^{*(\rm +)}_{f}\right]/2\omega.  
\label{atomic}
\ee
As was already mentioned in the introduction, the continuum--continuum
delay, $\tau_{cc}$ in Eq.~(\ref{atomicdelayparts}), for different
partial wave paths in hydrogen $s\rightarrow p\rightarrow l_f$ show
negligible dependence on the final angular momentum $l_f$ being $s$ or
$d$ wave \cite{Dahlstrom201353}.  As a starting point for this work we
verify that this holds true also in helium by extraction of the
continuum--continuum delay of a particular angular-momentum path as
\be
\tau_{cc}(l_f)=\tau_a(l_f)-\tau_W,
\label{taucc}
\ee 
where $\tau_W$ is computed for the intermediate $p$-wave for energy
$\epsilon_f=\sqrt{2(2q\omega+\epsilon_i)}$.  The result for
$l_f=(0,2)\equiv (s,d)$ is shown in Fig.~\ref{Hetaucc}, where indeed
no difference in $\tau_{cc}(l_f)$ is discernable between the two final
angular momentum states.

For angle-resolved delays, however, the relative phases of the $(+)$
and $(-)$ processes between different final partial wave angular
states is important because these different final states will
interfere.
This is clear from the explicit form of a momentum state
\be
\psi^{(-)}_{\k}(\r)
=k^{-1/2}
\sum\limits_{l=0}^{L_{\rm max}}
\sum\limits_{\mu=-l}^{l}
i^l e^{-i\delta_l}Y^*_{l\mu}(\hat\k)Y_{l\mu}(\hat\r)R_{kl}(r) \ ,
\label{partial}
\ee
where we have applied the ingoing boundary condition.  Formally
$L_{\rm max}$ extends to infinty in Eq.~(\ref{partial}), but in our
case with of two photon absorption from the $1s$ state in helium, the
momentum state can be truncated at $L_{\rm max}=2$ and $\mu=0$.
Using Eq.~(\ref{MqoOa}) and Eq.~(\ref{partial}) we now construct the
complex amplitude for angle-resolved photoelectron by 
 absorption of two photons $\omega_{2q-1}$ and $\omega$ 
%
%
\be 
\label{abs}
{\cal M}^{\rm (+)}_{\k_\f}=k_f^{-1/2}
\sum_{l_f=0,2}i^{-l_f}e^{i\delta_f}Y_{l_f,0}(\n_\f)
M(f,\omega,\omega_{2q-1},i)
\\ 
\ee 
and for absorption of one photon $\omega_{2q+1}$
followed by stimulated emission of $\omega$ 
\be 
\label{emi}
{\cal M}^{(\rm -)}_{\k_\f}=k_f^{-1/2}
\sum_{l_f=0,2}i^{-l_f}e^{i\delta_f}Y_{l_f,0}(\n_\f)
M(f,-\omega,\omega_{2q+1},i)
\\ 
\ee 
both leading to the same final state with photoelectron momentum,
$\k_\f=k_f\n_\f$ with
$k_f=\sqrt{2\epsilon_f}=\sqrt{2(2q\omega+\epsilon_i)}$. The factor
$k_f^{-1/2}$ comes from momentum normalization while the states in $M$
are normalized to energy \cite{Starace1982}.  The probability for
directed photoemission is proportional to
\be
S_{2q}(\k_\f)=2\left| {\cal
M}^{(\rm -)}_{\k_f}\exp[i(\phi_{2q+1}-\varphi)] + {\cal
M}^{(\rm +)}_{\k_f}\exp[i(\phi_{2q-1}+\varphi)] \right|^2,
\label{probq}
\ee
where we, again, write explicitly the phases of the fields so
that the field amplitudes, $E$, inside ${\cal M}$ (and $M$) are real.
Eq.~(\ref{oscillation}) and Eq.~(\ref{atomicdelay}) give the angle-resolved atomic delay
\be 
\tau_a(\k_\f)=\Delta\theta_{2q}(\k_\f)/2\omega= \arg[{\cal
M}^{(-)}_{\k_\f}{\cal M}^{*(+)}_{\k_\f}]/2\omega,  
\ee
where, in contrast to the angle-integrated expression \eref{atomic}, 
the interference of the two final partial waves depends on the
direction of the vector $\k_\f$. 
Results for the angle-resolved atomic delay are given in Sec.~\ref{sec:results}, 
where we show that the angular dependence of the time delay can be
easily interpreted using LOPT as a competition of the continuum--continuum transitions
$Ep\to E'd$ and $Ep\to E's$ driven the the IR absorption. As can be expected, this
competition may become particularly intense near the geometric node of 
the $d$-spherical wave. 

\begin{figure}[ht]
\bp{0.5\textwidth}
\includegraphics[width=\textwidth]{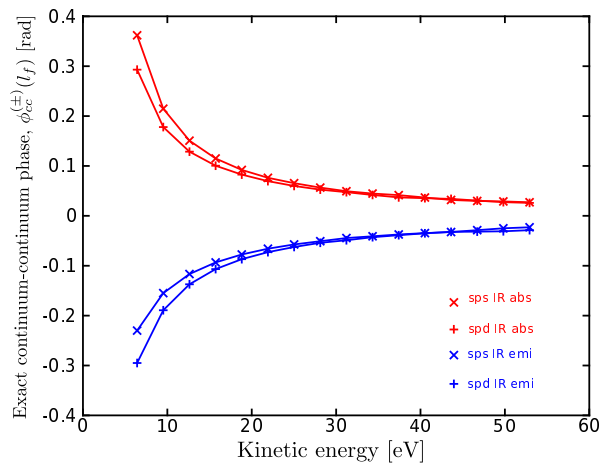}
\ep
\hs{-0.1\textwidth}
\bp{0.6\textwidth}
\caption{\label{Hephicc} Continuum--continuum phases, for absorption
and emission of an IR photon.  }
\ep
\end{figure}

In the spirit of \cn{Dahlstrom201353}, we now make a connection
between the continuum--continuum delay $\tau_{cc}$, and the
corresponding phase-shifts $\phi_{cc}^{(\pm)}$, of the two photon
matrix element.  We define ``exact'' $\phi_{cc}^{(+)}(\epsilon_fl_f)$
and $\phi_{cc}^{(-)}(\epsilon_fl_f)$, for absorption and emission of
an IR photon to the final state with angular momentum $l_f$ with
energy $\epsilon_f$ as
\ba
\phi_{cc}^{(+)}(\epsilon_f l_f)&=&\arg M^{(+)}_f -
\frac{\pi}{2}(l_f-2)-\delta_{2q-1}+\delta_f \nonumber \\
\phi_{cc}^{(-)}(\epsilon_f l_f)&=&\arg M^{(-)}_f -
\frac{\pi}{2}(l_f-2)-\delta_{2q+1}+\delta_f,
\label{phiccae}
\ea 
where $\delta_{2q-1}$ and $\delta_{2q+1}$ are the atomic scattering phases 
of the on-shell intermediate $p$-wave and $\delta_f$ is that of the final $s$ or $d$-wave.  
The result is presented in Fig.~\ref{Hephicc}, where
we observe that the CC-phases leading to different angular momentum final states 
differ for low kinetic energy electrons. 
Comparing with the exact calculation for hydrogen, Fig.~3 in
Ref.~\cite{Dahlstrom201353}, a similar level of discrepancy between
the $sps$ and $spd$ angular momentum paths is identified. This shows that
the deviations from the approximate continuum--continuum phases arise
already in hydrogen and that they do not require any additional
short-range interactions (such as correlation effects).
The question arises why the CC-delay of the $s$ and $d$-waves are identical, 
as was shown in Fig.~\ref{Hetaucc}, when the CC-phases are obviously different close to threshold.  
Closer inspection shows that the CC-phases of the $d$-wave are slightly below those 
of the $s$-wave in {\it both} aborption and emission processes by nearly the same amount, say 
$\phi_{cc}(\epsilon_f d)\approx\phi_{cc}(\epsilon_f s)-\xi(\epsilon_f)$.
When computing the CC-delay from the CC-phases one takes the difference of
emission and absorption processes, 
$\tau_{cc}(l_f)=[\phi_{cc}^{(-)}(l_f)-\phi_{cc}^{(+)}(l_f)]/2\omega$, 
so that this constant phase difference cancels. 

\subsection{TDSE approach}
\label{sec:theory:TDSE}

We solve the TDSE for a helium atom described within a SAE approximation
\begin{equation}
\label{TDSE}
i {\partial \Psi(\r) / \partial t}=
\left[\hat H_{\rm atom} + \hat H_{\rm int}(t)\right]
\Psi(\r) \ ,
\end{equation}
where $\hat H_{\rm atom}$ is the Hamiltonian of the field-free atom
with an effective one-electron potential \cite{Sarsa2004163}.  The
Hamiltonian $\hat H_{\rm int}(t)$ describes the interaction with the
external  field and is written in the velocity
gauge
\be
\label{gauge}
\hat H_{\rm int}(t) =
 {\bm A}(t)\cdot \hat{\bm p} \ \ , \ \ 
{\bm A(t)}=-\int_{0}^{t}{\bm E(t')}\ d t'.
\ee
As compared to the alternative length gauge, this form of the
interaction has a numerical advantage of a faster convergence.

The vector potential of the APT is modeled as the sum of 11 Gaussian
pulses of altering polarity shifted by a half of the IR period
$T=2\pi/\omega$ :
\be
\label{vectorGauss}
\hs{-1cm}
A_x(t) = \sum_{n=-5}^5 (-1)^n A_n \exp\left(
-2\ln2
{(t-nT/2)^2\over \tau_x^2}
\right)
\cos\Big[\omega_x(t-nT/2)\Big] \ .
\ee
The amplitude of each pulse is defined as
$$
A_n = A_0
\exp\left(-2\ln2
{(nT/2)^2\over \tau_T^2}\right),
$$
where $A_0$ is the vector potential amplitude related to the field
intensity $ I = (\omega^2/c^2) A_0^2$.  
The XUV central frequency is $\omega_x = 1.378$~au~$=37.5$~eV. 
%
The time constants $\tau_x=0.14$~fs and $\tau_T=4.83$~fs 
determine the length of an XUV pulse and the APT
train, respectively. The field intensity of the APT is chosen at
$5\times10^8~\rm W/cm^2$ and the XUV frequency  $\omega_x\simeq
25\omega$. 

The vector potential of the IR pulse is modeled by the cosine
squared envelope
\be
\label{vectorSin2}
A(t) = A_0 \cos^2
\left(
{\pi (t-\tau)\over 2\tau_{\rm IR}}
\right)
\cos[\omega(t-\tau)] \ ,
\ee
with an intensity of $3\times10^{11}$~W/cm$^2$ and pulse duration of
$\tau_{\rm IR} = 14.5$~fs.  The IR pulse is shifted relative to the
APT by a variable delay $0\le\tau\le 0.5T$.
A positive delay, $\tau>0$, corresponds to
the IR pulse being delayed with respect to the center of the XUV pulse train. 
Further, the laser photon energy is $\omega = 0.05841$~au~$=1.59$~eV,
which corresponds to a period of $T=2\pi/\omega=107$~au = 2.60~fs.
The laser pulse duration is $\tau = 5.58 T = 14.5$~fs.

\begin{figure}[htb!]
\centering

\hs{-10cm}
\epsfxsize=6cm
\epsffile{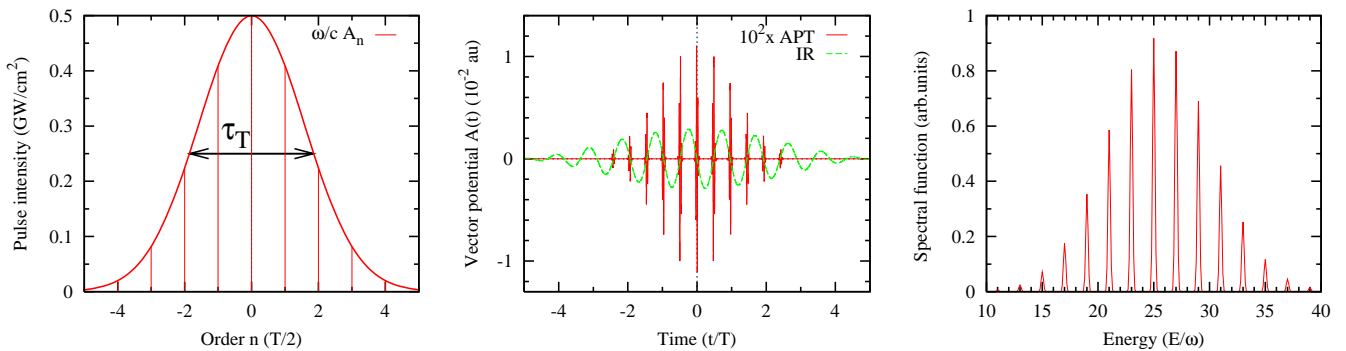}

 \caption{\label{Fig1} (Color online) Left: XUV pulses intensity
$(\omega^2/c^2)A_n^2$ (in GW/cm$^2$). The arrow visualizes
$\tau_T=1.86T=4.83$~fs Center: The vector potentials of the APT (red
solid line) and the IR pulse (blue dashed line). Right: spectral
representation of the vector-potential.   The XUV frequency
$\omega_x=37.5$~eV is chosen to match the 25st harmonic of
$\omega=1.59$~eV. The APT/IR delay $\tau=0$.  }

\end{figure}

\noindent 
The vector potential of the APT [Eq.~\eref{vectorGauss}] and the IR
pulse [Eq.~\eref{vectorSin2}] are visualized on the central panel of
\Fref{Fig1} along with the squared APT amplitudes
$(\omega^2/c^2)A_n^2$ (left) and the APT spectral content (right).

To solve the TDSE, we follow the strategy tested in our previous works
\cite{dstrong,PhysRevA.87.033407}.
The solution of the TDSE is presented as a partial wave series
\be
\Psi({\bm r},t)=
\sum\limits_{l=0}^{L_{\rm max}} 
f_{l}(r,t) Y_{l0}(\theta,\phi)
\label{basis}
\ee
with only zero momentum projections retained for the linearly
polarized light.  The radial part of the TDSE is discretized on the
grid with the stepsize $\delta r=0.05$ a.u. in a box of the size
$R_{\rm max}=2000$ a.u.  The number of partial waves in \Eref{basis}
was limited to $L_{\rm max}=5$ which ensured convergence in the
velocity gauge calculations.

\begin{figure}[htb!]

\epsfxsize=7cm
\epsffile{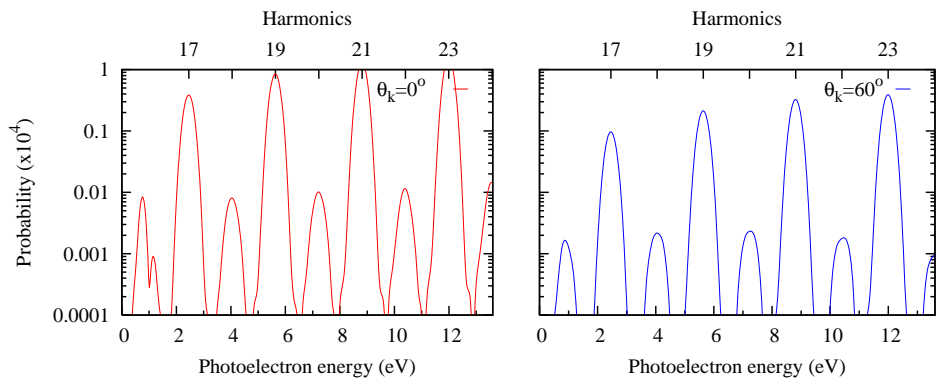}

 \caption{\label{Fig2} (Color online) The photoelectron spectra
 detected at the angles $\theta_k=0^\circ$ (left) and $\theta_k=60^\circ$
 (right)}

\end{figure}

Substitution of the expansion \eref{basis} into the TDSE gives a
system of coupled equations for the radial functions $f_{l\mu}(r,t)$,
describing evolution of the system in time.  To solve this system, we
use the matrix iteration method  \cite{velocity1}.  The
ionization amplitudes $a(\k)$ are obtained by projecting the solution
of the TDSE at the end of the laser pulse on the set of the ingoing
scattering states of the target Hamiltonian~\eref{partial}.  
Squared amplitudes $|a(\k)|^2$ give the photoelectron spectrum in
a given direction $\hat\k$ determined by the azimuthal angle $\theta_k$. 
Examples of such spectra in the $\hat \z$ direction $\theta_k=0$ and
$\theta_k=60^\circ$ are shown in \Fref{Fig2}

\begin{figure}[htb!]

\epsfxsize=6.5cm
\epsffile{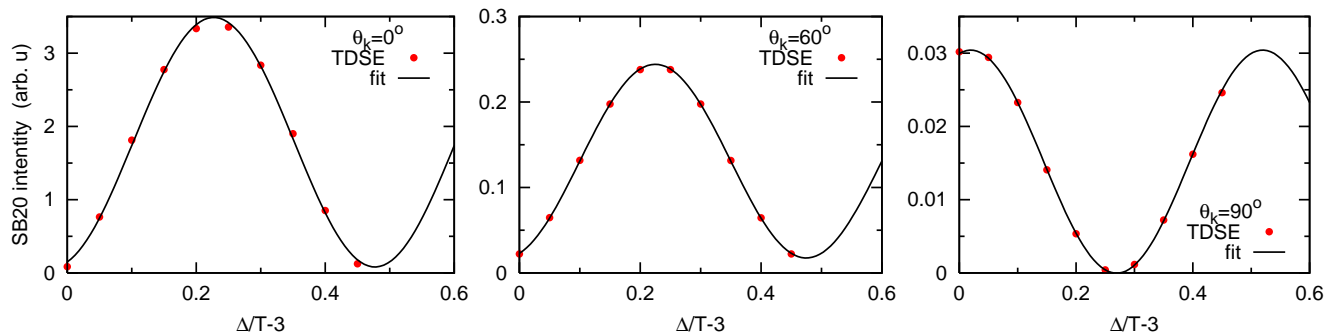}

 \caption{\label{Fig3} (Color online) The SB20 intensity oscillation
 as a function of the time delay $\Delta/T$ for the photoelectron
 detection angles $\theta_k=0^\circ$, $60^\circ$ and $90^\circ$ }
\end{figure}

This procedure is then repeated for various time delays $\tau$ between the APT and IR
fields and the SB intensity oscillations is fitted to \Eref{oscillation}  
for angle-resolved photoelectrons. 
After collecting the photoelectron spectra from TDSE in various directions, the
SB intensity oscillation with the variable time delay between the APT
and IR fields is fitted with the cosine function
\be
\label{Sfit}
S_{2q}({\k_f})=A+B\cos[2\omega\tau-C] 
\ee
using the non-linear Marquardt-Levenberg
algorithm. The quality of the fit is very good with the errors in all
three parameters not exceeding 1\%. Several examples of the fit for
the SB20 at the photoelectron detection angles $\theta_k=0^\circ$,
$60^\circ$ and $90^\circ$ are shown in \Fref{Fig3}.
%

\begin{figure}[htb!]

\hs{-0.5cm}
\epsfxsize=6.5cm
\epsffile{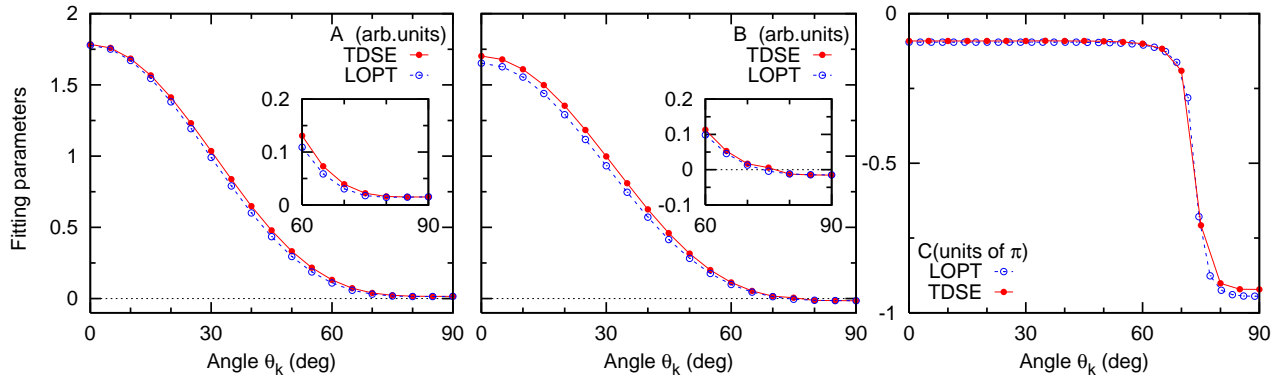}

\caption{\label{Fig4} (Color online) 
Angular dependence of the fitting parameters $A$, $B$, and $C$ for the
SB20. The TDSE and LOPT calculations are shown with the filled (red)
circles and open (blue) circles, respectively. The LOPT calculation is
normalized to the TDSE in the maximum of the $A$ parameter. The insets
show the variation of the $A$ and $B$ parameters near $90^\circ$.}
\end{figure}

\section{Results}
\label{sec:results}

In this section we compare the results from our frequency-domain and time-domain calculations.
The angular dependence of parameters $A$, $B$ and $C$ of
Eq.~(\ref{Sfit}) for the SB20 is shown in \Fref{Fig4} along with the
equivalent set of data from the LOPT calculation.  The LOPT
$A$-parameter is normalized to the same parameter in the TDSE
calculation. This normalization is then kept for the $B$-parameter.
The $C$ parameter is plotted on the absolute scale. All the three
parameters agree well between the TDSE and LOPT calculations.  We note
that even the change of sign of the $B$ parameter near $90^\circ$,
visible on the inset of the middle panel, is reproduced by both
calculations.

\begin{figure}[htb!]

\vs{4.7cm}
\hs{2.7cm}
\epsfxsize=6.5cm
\epsffile{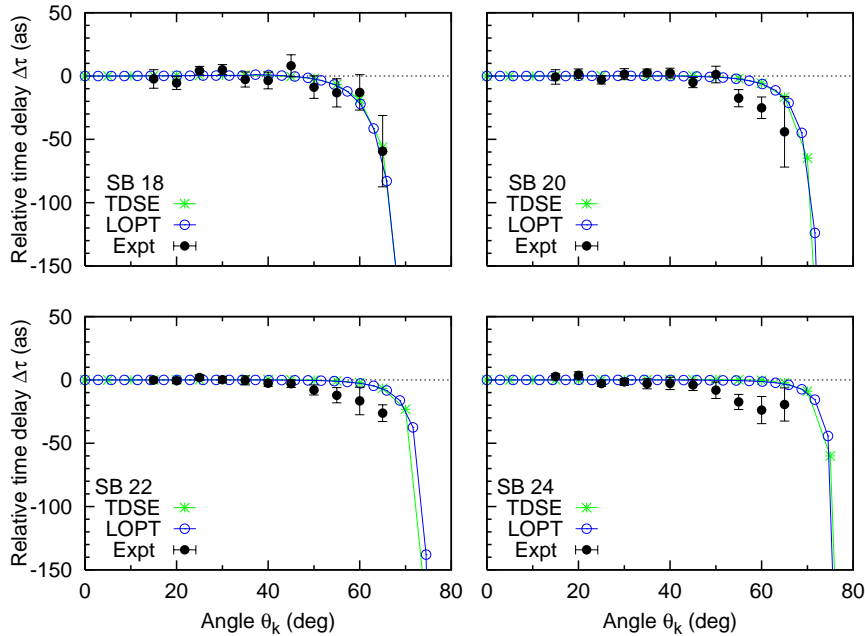}

\caption{\label{Fig5} (Color online) 
Variation of the time delay relative to the zero angle $\Delta\tau_a
= \tau_a(\theta_k) -\tau_a(0^\circ)$ for  SB 18, 20 (top)
and SB 22, 24 (bottom).  The TDSE results are shown with the 
(green) asterisks. The LOPT calculations are displayed with the (blue) open
circles.  The experimental data by \cn{2015arXiv150308966H} are
visualized with filled circles and error bars.}
\end{figure}

The group delay $\tau_{\rm GD}$ of the ATP is zero in our approach since we consider 
Fourier limited attosecond pulses by setting 
$\phi_{2q+1}=0$, for all integers $q$ in the frequency comb. 
Hence the parameter $C$ can be converted directly into the
atomic time delay as $\tau_a = C/2\omega$ according to
\Eref{oscillation}. 
The atomic time delay obtained in this fashion is given in
\Tref{Tab1} for the direction along the polarization of the field, 
which we refer to as the zero angle for photoemission.  
Again the agreement between the two calculations is good. 
To connect with Eq.~(\ref{atomicdelayparts}) we also show the
breakdown of the atomic delay into the Wigner time delay $\tau_W$,
which was computed separately by a one-photon RPAE program
\cite{PhysRevA.87.063404}, and the extracted continuum--continuum
delay $\tau_{cc}$. Finally we compare the extracted CC term with that
of earlier exact calculations in hydrogen \cite{Dahlstrom201353}.  The
discrepancy between the two CC quantities is reasonably small, less
than ten attoseconds. 
%
%
The variation of the atomic time delay relative to the zero angle
polarization direction $\Delta \tau = \tau_a(\theta_k)-\tau(0^\circ)$
is displayed in \Fref{Fig5} for SB 18 to 24. 

\begin{table}
\bp{12cm}
\caption{Atomic time delay $\tau_a$ and its various components 
$\tau_W$ and $\tau_{cc}$ in the $\hat\z$ direction for
  various side bands.
\label{Tab1}}
\bs

\begin{tabular}{cccccccc}
\hline\hline
SB &$n\omega$   &$E$ &\multicolumn{2}{c}{$\tau_a$ (as)}
 & $\tau_W$ (as) &\multicolumn{2}{c}{$\tau_{cc}$ (as)}\\
$n$ &eV &eV & TDSE & LOPT & RPAE & [1] & [2]\\
\hline\hline\\
18  & 27.9 &3.3 & -85     & -93 & 231 &-324 & -315 \\
20  & 31.0 &6.4 & -61     & -63 & 60  &-123 & -129 \\
22  & 34.1 &9.5 & -46     & -48 & 30  &-78  & -83  \\
24  & 37.2 &12.6& -37     & -37 & 16  &-53  & -57  \\
\hline\hline\\
\end{tabular}

{\footnotesize [1]  Atomic delay minus Wigner delay}\\
{\footnotesize [2] Fit to exact hydrogen calculation by 
Richard Ta\"ieb \cite{Dahlstrom201353}} \\
\ep
\end{table}

\begin{figure} [h]
\hs{-1cm}
\epsfxsize=6.cm
\epsffile{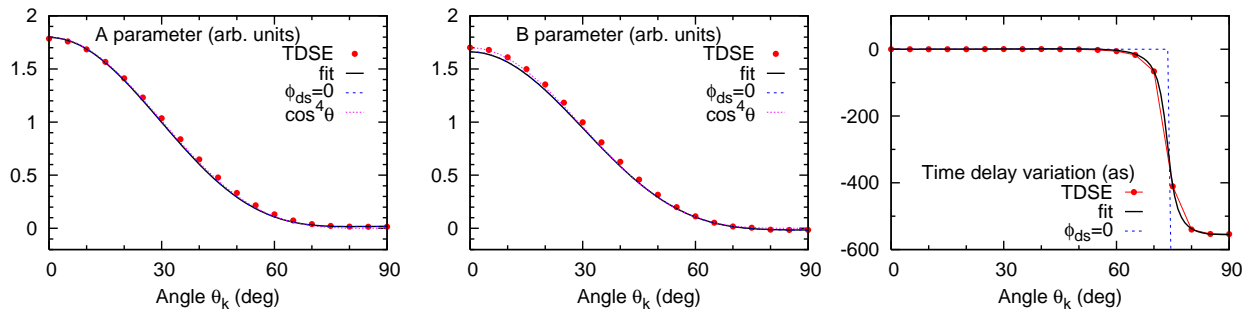}
\caption{Fitting the angular variation of the amplitude parameters $A$
  and $B$ and the time delay $\Delta\tau_a$ with \Eref{parameters} for
  SB 20. The (red) filled circles display the TDSE calculation whereas
  the thick solid line visualize the fit. The restricted set of
  fitting parameters with $\phi^\pm_{ds}=0$ is represented by the blue
  dashed line whereas the $\cos^4\theta$ fit to the $A$ and $B$
  parameters  is displayed with the (purple) dotted line.
\label{Fig6}}
\end{figure}

To highlight the competition between the $s$ and $d$-channels in 
two-photon ionization into the direction $\n_\f$, 
we parametrize the absorption and emission
amplitudes Eq.~\eref{abs} and Eq.~\eref{emi} in the form suggested in
\cn{2015arXiv150308966H}:
\be
\label{abem}
{\cal M}^{(\pm)}_{\k_\f}\propto
1 + c_{ds}^{(\pm)} e^{i\phi_{ds}^{(\pm)}}\sqrt{4\pi}Y_{20}(\n_\f) 
\ \ , \ \ 
\ee
where 
\be 
c_{ds}^{(\pm)}=|M_d^{(\pm)}/M_s^{(\pm)}|
\ee 
and
\be
\phi_{ds}^{(\pm)}=\arg[(i^{-2}e^{i\delta_d}M_d^{(\pm)})/(i^{-0}e^{i\delta_s}M_s^{(\pm))})
\ee 
are the amplitude
ratio and phase difference between the $d$ and $s$ partial wave
amplitudes of the absorption $(+)$ and emission $(-)$ processes,
respectively.
Using Eq.~(\ref{abem}), the $A$ and $B$ parameters and
the angular variation of the atomic time delay $\Delta\tau_a$ can be
presented as
\ba
\label{parameters}
A &\propto& 
\left|
1+c_{ds}^{(-)}e^{i\phi_{ds}^{(-)}}\sqrt{4\pi}Y_{20}(\n_\f)
\right|^2
+
\left|
1+c_{ds}^{(+)}e^{i\phi_{ds}^{(+)}}\sqrt{4\pi}Y_{20}(\n_\f)
\right|^2
\\ \nn
B &\propto& 
2{\rm Re} \left\{
\Big[1+c_{ds}^{(-)}e^{i\phi_{ds}^{(-)}}\sqrt{4\pi}Y_{20}(\n_\f)\Big]
\Big[1+c_{ds}^{(+)}e^{i\phi_{ds}^{(+)}}\sqrt{4\pi}Y_{20}(\n_\f)\Big]^*
\right\}
\\ \nn
\Delta\tau_a &=&
{1\over 2\omega} \arg
\left\{
\Big[1+c_{ds}^{(-)}e^{i\phi_{ds}^{(-)}}\sqrt{4\pi}Y_{20}(\n_\f)\Big]\Big/
\Big[1+c_{ds}^{(+)}e^{i\phi_{ds}^{(+)}}\sqrt{4\pi}Y_{20}(\n_\f)\Big]
\right\}.
\ea
Next, we fit the angular variation of the time delay $\Delta\tau_a$
from the TDSE calculation using the bottom line of \Eref{parameters}
with $c_{ds}^{(\pm)}$ and $\phi_{ds}^{(\pm)}$ as fitting parameters.
The result of this fitting procedure are illustrated in \Fref{Fig6}
for SB 20.  For comparison we also plot the case where we manually set
$\phi_{ds}^{(\pm)}=0$ so that all paths have the same phase.  While
this approximation has negligible effect on the amplitude
parameters $A$ and $B$, the atomic delay changes from a smooth step function,
which drops from $0$~as at $60^\circ$ to $-550$~as close to
$90^\circ$, to a discrete step that occurs close to 75$^\circ$.



It  follows from the soft photon approximation (SPA) \cite{Maquet2007} 
that the angular dependence of the $A$ and
$B$ parameters are simple $\cos^4\theta$ functions for an initial $s$-state,
\br
B\propto {\rm Re} 
\left[{\cal M}^{(-)}_{\k_\f}
{\cal M}^{*(+)}_{\k_\f}\right]
&\propto&
|J_1({\bm \alpha}_0\cdot\k_f)|^2
|\la f|z|i \ra|^2
\propto
\cos^4\theta \ .
\er
Here we made a linear approximation to the Bessel function as the
parameter ${\bm \alpha}_0={\bm F}_0/\omega^2$ is small in a weak IR
field.  This simple dependece fits very well both the $A$ and $B$
parameters. The only deviation occurs at large angles where the $B$
parameter becomes negative while $\cos^4\theta$ always remains
positive. However, the SPA predicts no angular
variation of the time delay. So, angular dependent time delay and
alterantion of sign of the $B$ parameter are both signs of breakdown
of the SPA. 

Numerical values of the $c_{ds}^{(\pm)}$ and $\phi^{(\pm)}_{ds}$
parameters for SB 20 and 22 are shown in \Tref{Tab2} along with the
LOPT calculation and the fully {\em ab initio} TDSE calculation
\cite{Argenti2014}. The latter TDSE calculation is not restricted by
the SAE and the two electrons in the He atom are treated on the equal
footing. Agreement between all the three calculations is fairly good.
We find that $\phi_{ds}^{(\pm)}$ is small and that it tends to
decrease with the energy of the photoelectron (side band order). This
is in qualitative agreement with the earlier work on the
atomic delay \cite{Dahlstrom201353}, where it was found that 
{\it no} phase difference was expected for sufficiently energetic electrons.  
Here, we study photoelectrons close to the ionization threshold and effects beyond the asymptotic approximation are at play.

\begin{table}[h]
\caption {Numerical values of the fitting parameters $c_{ds}^{(\pm)}$ and
  $\phi^{(\pm)}_{ds}$ for SB 20 and 22 from the present TDSE-SAE
  calculation, fully {\em ab initio} TDSE calculation by
  \cn{Argenti2014} and the LOPT calculation
\label{Tab2}}
\bs
\hs{2cm}
\bt{r|rrr|rrrrrrrrrrr}
& \mc{3}{|l}{SB 20~~ {\sc tdse} } 
& \mc{3}{|l}{SB 22~~ {\sc tdse} } \\
Parameter & {\sc sae} & \hs{-2mm}{\em ab initio} & {\sc lopt}  
          & {\sc sae} & \hs{-2mm}{\em ab initio} & {\sc lopt} \\
\hline\hline \\
$c_{ds}^{(+)}$    & 1.168 & 1.174 & 1.15  & 1.093 &1.098 & 1.08\\
$c_{ds}^{(-)}$    & 0.633 & 0.677 & 0.69  & 0.722 &0.685 & 0.73\\
$-\phi_{ds}^{(+)}$ & 0.090 & 0.082 & 0.061 & 0.043 &0.056 & 0.033\\
$-\phi_{ds}^{(-)}$ & 0.074 & 0.076 & 0.056 & 0.040 &0.047 & 0.031\\
\et
\end{table}

The amplitude ratio $c_{ds}^{(\pm)}$ is rather close to unity, which
means that the relative weight of the $d$ and $s$ channels in the
two-photon ionization process is approximately equal. This
demonstrates that Fano's propencity rule is not applicable for
transitions in the continuum. Although not shown in \Tref{Tab2}, the
LOPT calculation shows that the amplitude ratio for much higher
energies numerically approaches the kinematic limit of
$\sqrt{4/5}\approx0.89$. This high-energy limit indicates that the absorption and
emission processes become equal in magnitude and that they have no
relative phase difference. The Fano propencity rule does not hold for the second photon as the magnitude of the amplitude ratio is smaller than one.

\section{Conclusion}
\label{sec:conclusion}

In the present work we studied the angular variation of the atomic
time delay in the RABBIT measurement on the helium atom. We applied
the non-perturbative TDSE method along with the lowest order
perturbation theory with respect to the electron-photon
interaction. Our results are compared favourably with the recent
COLTRIMS measurement by \cn{2015arXiv150308966H}.

In the experimentally accessible angular range of 0 to $65^\circ$,
where the RABBITT signal is sufficiently strong, the angular variation
of the time delay is rather small.  Only SB18 displays a noticeable
angular variation of about 60~as. The time delay on other side bands
remain flat in this angular range. Given the rapid drop of the
magnitude $A$ and $B$ parameters in \Eref{oscillation} with the
detection angle as $\cos^4\theta$, the angular averaged time delay
$\bar \tau_a$ will be very close to that recorded in the polarization
direction of light at the zero degree angle $\tau_a(\theta=0)$ . This
allows to use the helium atom as a convenient standard both in the
angular specific streaking and angular averaged RABBITT time delay
measurements.
%

The partial wave analysis indicates that the $d$ and $s$ channels are
equally important in the two-photon ionization continuum both for the
angular variation of the magnitude $A$ and $B$ parameters as well as
the atomic time delay $\tau_a$ in the whole angular range. This is
contrary to the intuitive assumption that the $d$ wave normally
outweighs the $s$ wave and their competition becomes noticeable only
beyond the magic angle. The parametrization with the modulus ratio of
the $s$ and $d$ ionization amplitudes and their relative phase in the
absorption and emission channels provides a convenient tool to analyze
influence of various factors on the RABBITT signal. It also allows to
make a quantitative comparison between various
calculations. Unfortunately, the statistics of the experimental data
\cite{2015arXiv150308966H} is insufficient to extract these parameters
and to compare with theoretical predictions. We hope that this
statistics will improve in the future to facilitate such a
comparison. 

We also intend to apply this analysis to the angular variation of time
delay in heavier noble gas atoms, Ne and Ar, as well as in the
hydrogen molecule. The asymptotic properties of the two-photon
ionization amplitude  and the atomic time delay
 hold in this case as well provided there is a strongly
dominant single-photon transition $l_i\to\l$ from the initial state.
According to the propensity rule \cite{PhysRevA.32.617}, the dipole
transition with the increased momentum $l= l_i+1$ is usually dominant
unless the dominant channel passes through the Coopers minimum.  The
molecular case introduces an additional degree of freedom of
orientation of the inter-nuclear axis. Hence the physics of the
angular dependent time delay becomes significantly richer. This work
is in progress now \cite{2016arXiv160404938S}.

\ack The authors are grateful to Luca Argenti, \'Alvaro Himenez
Gal\'an, Fernando Mart\'{i}n, Sebastian Heuser, Claudio Cirelli and
Ursula Keller for many stimulating discussion. The group of Ursula
Keller at ETH Zurich kindly allowed us to use their experimental data.
The group of Fernando Mart\'{i}n at the UAM Madrid made for us
available their {\em ab initio} TDSE results.  
J.M.D. acknowledges support from the Swedish Research Council, 
Grant No.  2013-344 and 2014-3724. 
E.L. acknowledges support from the Swedish Research Council, 
Grant No. 2012-3668.
I.A. and A.S.K. acknowledge support by the Australian Research Council in the form of the Discovery grant DP120101805. I.A. acnowledges support from the
Institute for Basic Science, Gwangju, Republic of Korea.
J.M.D., E.L. and A.S.K. acknowledge the support of the Kavli Institute for Theoretical Physics (National Science Foundation under grant number NSF PHY11-25915).

\newpage

\section*{References}


\end{document}